\documentclass[runningheads]{llncs}
\usepackage{graphicx}
\usepackage{siunitx}
\usepackage{multirow}

\begin{document}
\title{An Optimization Framework for Processing and Transfer Learning for the Brain Tumor Segmentation}
\titlerunning{An Optimization Framework for BraTS 2023}
%
\author{Tianyi Ren\inst{*} \and Ethan Honey\inst{*} \and
Harshitha Rebala\inst{*} \and
Abhishek Sharma
\and Agamdeep Chopra
\and Mehmet Kurt
}

\authorrunning{T. Ren et al.} 
%

\institute{University of Washington, Seattle WA 98105, USA \\
\email{tr1@uw.edu}\\
}
\maketitle              


\def\thefootnote{*}\footnotetext{These authors contributed equally to this work}\def\thefootnote{\arabic{footnote}}

\begin{abstract}
Tumor segmentation from multi-modal brain MRI images is a challenging task due to the limited samples, high variance in shapes and uneven distribution of tumor morphology. The performance of automated medical image segmentation has been significant improvement by the recent advances in deep learning.
However, the model predictions have not yet reached the desired level for clinical use in terms of accuracy and generalizability.
In order to address the distinct problems presented in Challenges 1, 2, and 3 of BraTS 2023, we have constructed an optimization framework based on a 3D U-Net model for brain tumor segmentation. This framework incorporates a range of techniques, including various pre-processing and post-processing techniques, and transfer learning.
On the validation datasets, this multi-modality brain tumor segmentation framework achieves an average lesion-wise Dice score of 0.79, 0.72, 0.74 on Challenges 1, 2, 3 respectively.

\end{abstract}

\section{Introduction}

Brain tumors are a deadly type of cancer that have long-term impacts on a patient's health \cite{adewole2023brain}. Treatments of surgery, radiation therapy, and drug therapy are pursued depending on the type and condition of the patient’s brain tumor \cite{baid2021rsnaasnrmiccai}. These insights into the brain tumor are often obtained through Magnetic Resonance Imaging (MRI) scans, which require experienced radiologists to manually segment tumor sub-regions \cite{baid2021rsnaasnrmiccai}. This is a long process that is unscalable to the needs of all patients. Thus, the recent growth of deep learning technologies holds promise to provide a reliable and automated solution to segmentation to save time and help medical professionals with this process \cite{Luu2022}.

The BraTS Challenge encourages advancement in the field of tumor segmentation of brain MRI images by providing challenges geared toward specific brain tumors. Challenge 1 focuses on adult glioblastoma, which is one of the most common tumors. Over the past 20 years, there has been little to no change in the average patient prognosis of 14 months \cite{baid2021rsnaasnrmiccai,6975210,GLI-3rdSource}. This highlights the need of faster yet accurate tumor segmentation for clinical treatments and experiments. 

Challenge 2 is centered on glioblastoma in Sub-Saharan Africa, where cases and mortality rates are still rising \cite{adewole2023brain}. These MRI scans are often lower in resolution and contrast and thus require tailored segmentation models to bridge this gap \cite{adewole2023brain}. The use of an automatic and accurate segmentation would improve healthcare access to more patients in Sub-Saharan Africa since most of the current tumor segmentation analysis is limited to high-resource urban areas.

The focus of Challenge 3 is intracranial meningioma. Since imaging is one of the most common ways to diagnose meningioma, automated segmentation would be pivotal in assisting surgical and radiotherapy planning by approximating tumor volume and sub-regions \cite{labella2023asnrmiccai}.

For all challenges, each model is evaluated by computing the Dice similarity coefficient and Hausdorff distance (95\%) (HD95) between each region of interest in the ground truth and the corresponding prediction. In previous years, these metrics were calculated over the whole brain volume; however, this year they are computed lesion-wise, giving equal weight to each tumor lesion regardless of its size. There are also extra penalties for false negative (FN) and false positive (FP) predictions. 
Consequently, we investigated new pre- and post-processing techniques alongside a range of loss functions to maximize our performance in the new lesion-wise metrics, particularly to minimize the number of FP predictions.

Various deep learning architectures, including classical convolutional neural networks, widely used U-Net \cite{cciccek20163d}, recently popular vision transformers \cite{dosovitskiy2020image} and swin transformer \cite{hatamizadeh2021swin}, have been applied in automating brain tumor segmentation and classification tasks. The state-of-the-art models in brain tumor segmentation are based on the encoder-decoder architectures such as U-Net \cite{isensee2021nnu} and its variants. For instance, Luu et. al \cite{Luu2022} modified the  nnU-Net model by adding an axial attention in the decoder. Futrega et. al \cite{futrega2021optimized} optimized the U-Net model by adding foreground voxels to the input data, increasing the encoder depth and convolutional filters. Siddiquee et. al \cite{siddiquee2021redundancy} applied adaptive ensembling to minimize redundancy under perturbations.

Consequently, we propose deep learning framework for automatic brain tumor segmentation. This frameworkaims to solve various problems in multiple tasks such as model prediction using of U-Net, image pre-processing techniques given a certain criteria, post-processing of the model prediction, and transfer learning training strategies for low-quality images.

\section{Methods}
We employed the same overall pipeline for Challenges 1 and 3, which we will outline in Subsections 2.1--2.6.
\subsection{Data and pre-processing} \label{sub-sec:pre-proc}
The BraTS mpMRI scans for each patient comprise of a native T1, post-contrast T1-weighted, T2-weighted, and T2 Fluid Attenuated Inversion Recovery (T2-FLAIR) volumes. The dataset was preprocessed by applying DICOM to NIfTI file conversions, and coregistration to the same SRI24 anatomical template. The preprocessing techniques of resampling to a uniform isotropic resolution (1mm3) and skull-stripping to maintain anonymity were also done \cite{baid2021rsnaasnrmiccai,adewole2023brain,labella2023asnrmiccai}. In addition to the standardized preprocessing applied to all BraTS mpMRI scans by the BraTS Challenge organizers, our team experimented with further processing techniques to improve the quality and comparability of the dataset. 

\subsubsection{Z-Score Normalization}
The Z-score normalization technique was used to address the varying intensity distributions across the dataset as seen in \cite{Luu2022}. Z-score normalization can be computed by calculating the mean and standard deviation of voxels and then transforming the value of the voxel by first subtracting the mean and dividing it by the standard deviation. This technique standardizes the intensity values of each voxel and allows for better direct comparisons between the different patient data by removing outliers.
\subsubsection{Rescaling Voxel Intensities}
Segmentation relies on the ability to identify important features and structures, so we aimed to improve the visibility of such features by rescaling voxel intensities. The voxel intensity percentiles were calculated and we defined the 2nd and 98th percentiles to be the range of voxel intensity to be stretched to cover the entire intensity range. This technique increases the contrast and provides more insight into the subtle features in the data. 
\subsubsection{Histogram Contrast Matching}
 Histogram contrast matching helps in better aligning the intensity distributions between the image contrasts. This technique transforms the source data's voxel values in a manner that ensures that the new histogram matches a reference data sample. 


\subsection{Model architecture} \label{sub-sec:model}
\begin{figure}
\centering
\includegraphics[width=\textwidth]{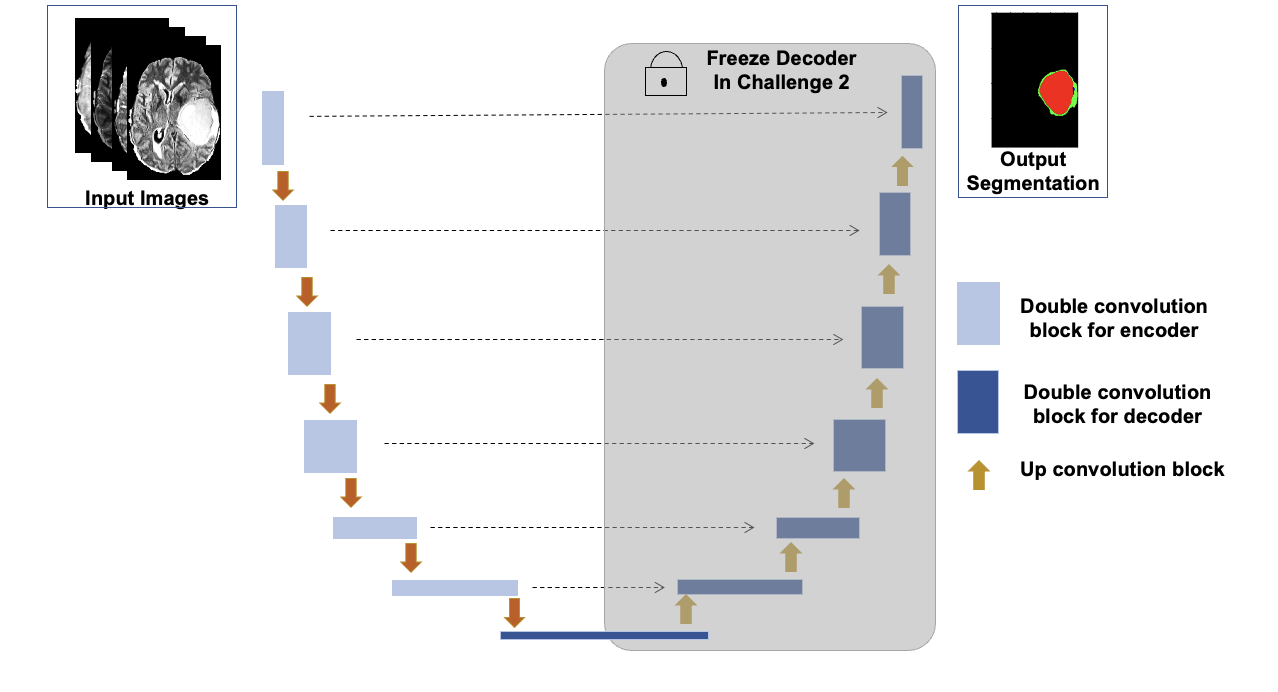}
\caption{(a) Baseline U-Net model \cite{futrega2021optimized} } \label{fig1}
\end{figure}

We used the optimized U-Net \cite{futrega2021optimized} as our baseline model for challenge 1 and 3 which can be seen in Figure~\ref{fig1}a. 
We used Challenge 1's model, froze its decoder layers as seen in Figure~\ref{fig1}a and then continued training on Challenge 2 dataset for transfer learning.

\subsection{Loss Functions} \label{sub-sec:loss}
The selection of loss functions plays an important role in guiding accurate segmentation results, so we explored various loss functions to determine the most effective ones. 

We first investigated the Mean Squared Error (MSE) loss to enhance the model’s accuracy. MSE provides the average squared difference between the predicted and ground truth voxels and therefore is useful in attaining accuracy at the voxel-level by heavily penalizing large deviations. Next, we explored Cross-Entropy (CE) loss, which  measures the dissimilarity between predicted and ground truth labels \cite{jadon2020survey} and penalizes false negatives and false positives. Another contender was Dice loss (Dice) since it acts as a strong indicator of similarity by calculating the overlap between the predicted and ground truth regions \cite{jadon2020survey}. This made it a natural fit for handling class imbalance and enhancing the model’s ability to accurately segment the regions of interest. We also considered Focal loss (Focal) to enhance the model's focus on segmenting difficult regions. Focal loss focuses learning by assigning a higher weight to challenging data \cite{jadon2020survey}. In order to reinforce correct boundary deformation and overcome intensity dependence, we introduced an Edge loss (Edge) component, as described in \cite{abderezaei20223d}. Edges were obtained by computing and normalizing the magnitude of the extracted intensity gradients. The Edge loss was calculated by comparing the edges of the predicted image to the target image edges using MSE.


\subsection{Training details} \label{sub-sec:training}

The ground truth segmentations provided by the BraTS organizers are in the following format: each voxel in the volume has one of the following labels: 1 for necrotic tumor core (NCR), 2 for peritumoral edematous/invaded tissue (ED), 3 for enhancing tumor (ET) and 0 for background voxels (no tumor).
However, the evaluation is performed with partially overlapping regions. These consist of enhancing tumor; tumor core (TC), the union of ET and NCR; and whole tumor (WT), the union of ET, NCR and ED.
Thus, from the ground truth labels we produce three channels that correspond to the partially overlapping regions ET, TC and WT. We compute the loss between these channels and our model output, so that the model output for a given voxel is a set of probabilities that the voxel lies in each of these overlapping regions. For our loss functions described above, we compute the loss between the ground truth and model output for each channel separately and then average them.

We used the Adam optimizer with no weight decay and learning rate equal to \num{6e-5}. Additionally, we implemented exponential decay of the learning rate.
We divided the official training dataset into a 90:10 split, holding 10\% of the dataset for validation rather than training. After every epoch, we iterate through this validation hold, calculating the Dice and HD95 score between our prediction and the ground truth for the regions of ET, TC, and WT and then averaging these scores. We then average this over each subject in the validation hold and save the model checkpoints with the highest such score.

\subsection{Inference} \label{sub-sec:inf}


As mentioned, we construct the training of our model such that it outputs a probability for each voxel being in each of the overlapping regions ET, TC and WT. However, the final output of our model must associate to each voxel a single label that is either 1 (NCR), 2 (ED), 3 (ET) or 0 (no tumor). Thus, we implement a thresholding strategy as in \cite{futrega2021optimized} to convert the probabilities for overlapping regions to actual predictions for the disjoint labels.

Firstly, consider the WT probability for a given voxel. If it is not large enough (less than 0.45), we do not consider the voxel to be in the tumor and set its label to 0 (no tumor), otherwise, we consider the voxel in the tumor.
Next, consider the TC probability for the voxel. If it is not large enough (less than 0.4), we do not consider the voxel to be in TC and set its label to 2 (ED), otherwise, we consider the voxel to be in TC.
Finally, consider the ET probability for the voxel. If it is not large enough (less than 0.45), we do not consider the voxel to be in ET and set its label to 1 (NCR), otherwise, we consider the voxel as ET and set its label to 3.

\subsection{Post-processing} \label{sub-sec:post-proc}

To prepare our final predictions, we perform the following strategy. We identify all connected components of ET voxels and remove any that have a volume less than or equal to 50 voxels, by relabelling the prediction labels to 0 for these voxels. (We call this removing `dust'.) Then we examine the TC voxels (ET + NCR) and check if any of the ET voxels that we removed created any holes\footnote{By holes, we mean a connected component of background voxels that is fully enclosed by the corresponding tumor region voxels.} in this region; if so, we change these voxels to be NCR (label 1). Next we remove any dust that still exists in the TC voxels, before examining the WT voxels (ET + NCR + ED) and filling any holes created by the removal of TC dust by changing these voxels to be ED (label 2). Finally, we remove any dust that still exists in the WT voxels.


\subsection{Transfer Learning for Challenge 2}

We applied the same pre- and post-processing techniques for the Challenge 2 data as we used for Challenges 1 and 3, but opted for a different training approach. One major hurdle for Challenge 2 was the limited training dataset of 60 patients which made training a full-scale, accurate model difficult. Therefore, we decided to implement transfer learning \cite{9077067} which uses a pre-trained model that already understands the important features of the task and then freezes or adds new layers to adapt the model to a new yet similar task. In our case, we utilized our model from Challenge 1 (also focused on glioma) and continued training on Challenge 2 data while experimenting with different frozen layers. Different strategies has been investigated such as freezing the encoder, decoder, and no layers. We also froze the middle layers, which included convolutional layers 5 to 7 in the encoder and layers 6 to 4 in the decoder. 


Freezing in this context of transfer learning refers to the designated layer’s weights and biases being prevented from updating during training. Instead, the layers retain the values and thus the learned features from the pre-trained model. In the context of Challenge 2, we considered freezing to be the optimal solution to prevent overfitting to the small dataset and maintain feature retention from Challenge 1’s model, which was trained on higher-resolution data. For our final model, we froze the decoder, as illustrated by the grey box as seen in Figure~\ref{fig1}a.

\section{Results}

\subsection{Pre-processing}
To determine the most effective pre-processing technique, we tested Z-score normalization, rescaling voxel intensities, and histogram contrast matching. We ran an experiment by randomly selecting a subset of 100 glioblastoma patients and applying different pre-processing techniques before training. The data was processed in the following manners for each trial: (1) no additional pre-processing (2) Z-score normalization, (3) Z-score \& rescaling, and (4) Z-Score \& rescaling \& histogram-matching. The model was then validated on a random sample of 30 glioblastoma patients, and we recorded the overall average Dice score for the ED, ET and NCR across all patients. The model's features were kept consistent throughout these experiments and the pre-processing experiment results can be found in Table \ref{tab:pre-process}. As a result, we rescale the voxel intensity after Z-Score normalization as the preprocessing protocol for all the 3 challenges.

\begin{table}[]
\centering
\caption{Average Dice across tumor sub-regions for different pre-processing techniques}\label{tab:pre-process}
\begin{tabular}{|l|c|}
\hline
\textbf{Preprocessing Techniques} & \multicolumn{1}{l|}{\textbf{Dice}} \\ \hline
No Additional Preprocessing      & 0.7539                                                             \\
Z-score normalization            & 0.7432                                                             \\
Z-Score \& Rescaling            &\textbf{0.8153}                                                           \\
Z-Score \& Rescaling \& Histogram-matching    
            &  0.8017                                                      \\ \hline
\end{tabular}
\end{table}

\subsection{Loss functions}
Different loss functions were also been investigated, including MSE, CE, Focal, Dice Loss, and a previously developed loss function Edge loss \cite{abderezaei20223d}. We examined three distinct combinations for our loss functions: (1) MSE, CE, and Edge. We obtained their respective weights of 0.25, 0.0044 and 0.00015 through hyperparameter optimization using Optuna \cite{optuna_2019}.
(2) Dice, Focal, and Edge. The weights were set to sensible values of 1, 1, and 0.05, respectively.
(3)	Similar to combination 2, Dice, Focal, and Edge were included in the compound loss. However, the weight edge loss is set to 0.005. The weights of Dice and Focal are still 1.

The weight of Dice and Focal losses were set to 1, as they were on a similar scale, and the weights edge loss was set to $0.05$ or $0.005$, based on the relative scale of the output of the Edge loss function in comparison to Dice and Focal losses. Models were trained on the entire Challenge 1 dataset with these different loss function combinations and evaluated during the validation phase. The results can be seen in Table \ref{tab:GLI_Loss}.

\begin{table}[]
\centering
\caption{Lesion-wise Dice and HD95 scores for Challenge 1 Loss Function Combinations}\label{tab:GLI_Loss}
\begin{tabular}{l|ccc|ccc|}
\cline{2-7}
                                                & \multicolumn{3}{c|}{\textbf{Dice}}                                    & \multicolumn{3}{c|}{\textbf{HD95}}                            \\ \hline
\multicolumn{1}{|l|}{\textbf{Loss functions}} & \multicolumn{1}{c|}{\textbf{ET}} & \multicolumn{1}{c|}{\textbf{TC}} & \textbf{WT} & \multicolumn{1}{c|}{\textbf{ET}} & \multicolumn{1}{c|}{\textbf{TC}} & \textbf{WT} \\ \hline
\multicolumn{1}{|l|}{MSE + CE + Edge}    & \multicolumn{1}{c|}{\textbf{0.7696}}      & \multicolumn{1}{c|}{\textbf{0.7980}}      & 0.8024      & \multicolumn{1}{c|}{\textbf{22.40}}       & \multicolumn{1}{c|}{\textbf{28.59}}       & 46.06       \\ \cline{2-7} 
\multicolumn{1}{|l|}{Dice + Focal + 0.05*Edge}             & \multicolumn{1}{c|}{0.7413}      & \multicolumn{1}{c|}{0.7796}      & \textbf{0.8296}     & \multicolumn{1}{c|}{39.26}       & \multicolumn{1}{c|}{34.59}       & \textbf{32.30}       \\ \cline{2-7} 
\multicolumn{1}{|l|}{Dice + Focal + 0.005*Edge}    & \multicolumn{1}{c|}{0.7523}      & \multicolumn{1}{c|}{0.7742}      & 0.8257      & \multicolumn{1}{c|}{30.61}       & \multicolumn{1}{c|}{30.69}       & 35.72       \\ \hline
\end{tabular}
\end{table}

Analyzing the results, combination 1 performed the best on ET and TC as per the Dice and HD95 evaluation metrics from the validation phase. However, it was important to note its HD95 score for WT was significantly larger compared to the other experiments. Combination 3's Dice scores were on par with combination 2 and its average HD95 score was lower. Therefore, combination 1 and 3 became our top contenders and we further tested on Challenge 3, meningioma. This was done similarly to the previous experiment by training with the different loss function combinations on the same datasets. The results in table \ref{tab:MEN_Loss} show that using a compound loss function including Dice loss, Focal loss, and Edge loss gives us better results in Dice socre and HD95 score which is consistent with challenge 1.

\begin{table}[]
\centering
\caption{Lesion-wise Dice and HD95 scores for Challenge 3 Loss Function Combinations}\label{tab:MEN_Loss}
\begin{tabular}{l|ccc|ccc|}
\cline{2-7}
                                                & \multicolumn{3}{c|}{\textbf{Dice}}                                    & \multicolumn{3}{c|}{\textbf{HD95}}                            \\ \hline
\multicolumn{1}{|l|}{\textbf{Loss functions}} & \multicolumn{1}{c|}{\textbf{ET}} & \multicolumn{1}{c|}{\textbf{TC}} & \textbf{WT} & \multicolumn{1}{c|}{\textbf{ET}} & \multicolumn{1}{c|}{\textbf{TC}} & \textbf{WT} \\ \hline
\multicolumn{1}{|l|}{MSE + CE + Edge}    & \multicolumn{1}{c|}{0.7494}      & \multicolumn{1}{c|}{0.7136}      & 0.7240      & \multicolumn{1}{c|}{49.52}       & \multicolumn{1}{c|}{65.31}       & 59.39       \\
\multicolumn{1}{|l|}{Dice + Focal + 0.005*Edge}    & \multicolumn{1}{c|}{\textbf{0.7602}}      & \multicolumn{1}{c|}{\textbf{0.7565}}      & \textbf{0.7322}      & \multicolumn{1}{c|}{\textbf{43.90}}       & \multicolumn{1}{c|}{\textbf{44.26}}       & \textbf{55.99}       \\ \hline
\end{tabular}
\end{table}

\subsection{Post-processing}

Our models' results were predicting numerous small connected components that were being classified as false positives (FPs) under the new lesion-wise evaluation metrics and decreasing our scores. We experimented with post-processing strategies to address this.

Our main idea was to find any small connected components (of any overlapping region) and reset their labels to 0, referred to as removing `dust' by connected-components-3d \cite{Silversmith_cc3d_Connected_components_2021}, a Python package employed in the official evaluation code that we also utilized in our pipeline. We removed any components of 50 voxels or fewer, mirroring how the official evaluation code ignored ground truth lesions of this size. We also experimented with replacing the dust with the modal label of its neighboring voxels, which took a lot more computation time for very little gain.

The removal of dust from ET and TC created small holes in the predictions of TC and WT respectively. To resolve this, any holes that were created in TC were relabeled as NCR, and holes in WT as ED. This adjustment saw a very marginal increase in performance but definitely seemed more robust. The comparison of our final post-processing technique with no post-processing is shown in Table \ref{tab:post-proc}.

\begin{table}[]
\centering
\caption{Lesion-wise Dice and HD95 scores with and without post-processing}
\label{tab:post-proc}
\begin{tabular}{|l|rrr|rrr|}
\hline
\multicolumn{1}{|c|}{\multirow{2}{*}{\textbf{Post-processing strategy}}} & \multicolumn{3}{c|}{\textbf{Dice}}                                                                     & \multicolumn{3}{c|}{\textbf{HD95}}                                                                     \\ \cline{2-7} 
\multicolumn{1}{|c|}{}                                                   & \multicolumn{1}{l|}{\textbf{ET}} & \multicolumn{1}{l|}{\textbf{TC}} & \multicolumn{1}{l|}{\textbf{WT}} & \multicolumn{1}{l|}{\textbf{ET}} & \multicolumn{1}{l|}{\textbf{TC}} & \multicolumn{1}{l|}{\textbf{WT}} \\ \hline
No post-processing                                                                     & \multicolumn{1}{r|}{0.6337}      & \multicolumn{1}{r|}{0.7024}      & 0.5722                           & \multicolumn{1}{r|}{84.31}       & \multicolumn{1}{r|}{57.57}       & 138.30                           \\
Removing dust and filling holes                                          & \multicolumn{1}{r|}{\textbf{0.7130}}      & \multicolumn{1}{r|}{\textbf{0.7672}}      & \textbf{0.8090}                           & \multicolumn{1}{r|}{\textbf{43.91}}       & \multicolumn{1}{r|}{\textbf{26.76}}       & \textbf{40.40}                            \\ \hline
\end{tabular}
\end{table}

\subsection{Transfer Learning for Challenge 2}
Finally, we conducted an experiment on the transfer learning approach to determine which model if  pre-trained on challenge 1 data can help improve segmentation performance for challenge 2. We use two baselines in this experiments: 1) baseline 1: model trained on challenge 2 data only (challenge 2 model), 2) baseline 2: model trained on challenge 1 data only (challenge 1 model).
We also investigated which layers in the model contribute to improvements in performance.
Thus, we selected our pre-trained challenge 1 model trained with the loss functions of Dice, Focal, and Edge Loss, and continued training the model on challenge 2's dataset using the same loss functions and  weights, but with specific layers frozen. The results evaluated during the validation phase show that freezing the decoder gives us the best results (Table \ref{tab:frozen_layers}).

\begin{table}[]
\centering
\caption{Lesion-wise scores for challenge 2 Freezing Layers Experiment}
\label{tab:frozen_layers}
\begin{tabular}{|l|llll|llll|}
\hline
\multicolumn{1}{|c|}{\multirow{2}{*}{\textbf{Experiments}}} & \multicolumn{4}{c|}{\textbf{Dice}}                                                & \multicolumn{4}{c|}{\textbf{HD95}}                                                \\ \cline{2-9} 
\multicolumn{1}{|c|}{}                                      & \multicolumn{1}{l|}{\textbf{ET}} & \multicolumn{1}{l|}{\textbf{TC}} & \multicolumn{1}{l|}{\textbf{WT}} & \textbf{mean} & \multicolumn{1}{l|}{\textbf{ET}} & \multicolumn{1}{l|}{\textbf{TC}} & \multicolumn{1}{l|}{\textbf{WT}} & \textbf{mean} \\ \hline

Challenge 2 model       
& \multicolumn{1}{l|}{0.6524}      & \multicolumn{1}{l|}{0.6925}      & \multicolumn{1}{l|}{0.5972} & 0.6474     & \multicolumn{1}{l|}{81.96}       & \multicolumn{1}{l|}{70.49}       & \multicolumn{1}{l|}{120.81} & 91.09     \\

Challenge 1 model                                          & \multicolumn{1}{l|}{0.7188}      & \multicolumn{1}{l|}{0.7424}      & \multicolumn{1}{l|}{0.6433} & 0.7015      & \multicolumn{1}{l|}{\textbf{52.35}}       & \multicolumn{1}{l|}{\textbf{48.16}}       & \multicolumn{1}{l|}{101.58} & \textbf{67.36}      \\

Freeze encoder layers                                              & \multicolumn{1}{l|}{0.6745}      & \multicolumn{1}{l|}{0.7143}      & \multicolumn{1}{l|}{0.5810} & 0.6566     & \multicolumn{1}{l|}{80.40}       & \multicolumn{1}{l|}{69.80}       & \multicolumn{1}{l|}{142.55} & 97.58      \\
Freeze middle layers                                              & \multicolumn{1}{l|}{\textbf{0.7392}}      & \multicolumn{1}{l|}{\textbf{0.7533}}      & \multicolumn{1}{l|}{0.6484} & 0.7136     & \multicolumn{1}{l|}{57.27}       & \multicolumn{1}{l|}{55.10}       & \multicolumn{1}{l|}{114.41} & 75.59      \\
Freeze decoder layers                                              & \multicolumn{1}{l|}{0.7200}      & \multicolumn{1}{l|}{0.7358}      & \multicolumn{1}{l|}{\textbf{0.7039}}  & \textbf{0.7199}     & \multicolumn{1}{l|}{59.21}       & \multicolumn{1}{l|}{60.21}       & \multicolumn{1}{l|}{\textbf{86.48 }} & 68.63      \\
Freeze nothing                                              & \multicolumn{1}{l|}{0.7159}      & \multicolumn{1}{l|}{0.7214}      & \multicolumn{1}{l|}{0.6764} & 0.7046      & \multicolumn{1}{l|}{68.72}       & \multicolumn{1}{l|}{67.43}       & \multicolumn{1}{l|}{103.55} & 79.90      \\ \hline
\end{tabular}
\end{table}

\section{Discussion}

The introduction of lesion-wise evaluation criteria pushed us to rethink our approach to automatic brain tumor segmentation. We systematically implemented and compared different strategies at different stages of the model training. For pre-processing, we wanted to ensure that our model has improved visibility into the four image modalities to identify the tumor sub-regions correctly. Our experiments with pre-processing techniques revealed that the best Dice score was achieved by applying both Z-score normalization and rescaling voxel intensities as seen in Table \ref{tab:pre-process}. As a result,  we decided to move forward with these two techniques for all the challenges. For model training, we investigated several combinations of loss functions. We first ran an experiment on challenge 1 as seen in Table \ref{tab:GLI_Loss} and then tested our top two combinations on challenge 3. Examining the challenge 3 results from Table \ref{tab:MEN_Loss}, The results shows that a compound loss function includes Dice, Focal, and Edge loss combination with the respective weights of 1, 1, $0.05$. obtained the best scores across all evaluation metrics. When computing the loss between the three output channels of our model and three corresponding channels of ground truth segmentations, there were two options for how to prepare the ground truth channels. They could be the disjoint labels NCR, ED, and ET, which is the format in which they are provided, or the partially overlapping regions WT, TC, and ET. We experimented with both options and observed marginal differences in the results. As a result, we picked the overlapped regions for training Since the official BraTS evaluation metrics are on the overlapping regions.



Also, we performed a rudimentary hyperparameter search using Optuna(A hyperparameter optimization framework)\cite{optuna_2019} to find the thresholds for converting model output probabilities to actual predictions, so that each voxel had exactly one label assigned to it. The results show that that any thresholds below 0.9 worked reasonably well, since the model's output probabilities were usually around 0.9 for voxels that it believed to be in the tumor region of interest. Thus, we chose thresholds based on \cite{futrega2021optimized}.

As mentioned before, decrease in performance was observed if the model was evaluated using the new lesion-wise metrics when compared to the `legacy' scores of BraTS challenges from previous years. This is because the extra penalties that false positive (FP) components receive. We investigated several approaches to overcome this problem, The best solution we found was in post-processing, as we see from Table \ref{tab:post-proc}, our final post-processing strategy improved overall performance in both Dice and HD95 scores in each of the partially overlapping regions. However, this post-processing strategy did decrease scores for certain subjects by removing small components in our prediction that did match up with the ground truth.


To understand why the so-called `dust' was being predicted by our models in the first place, we further analyzed the ground truth data (for challenge 1). We found that the ground truth segmentations themselves contain a large amount of dust, which is likely why the model was learning to predict this. However, as mentioned, the official evaluation code ignored any lesions that are 50 voxels or fewer from its calculations. Thus, we tried applying the dust-removal strategy as a pre-processing technique on the training data, to remove such lesions from the ground truth. For each region, we identify lesions using the same strategy as the official evaluation code (by considering separate connected components whose dilations overlap with each other as one lesion).

We implemented this for the training of our final model for Challenge 1, however, this still predicted a lot of FP components that we had to remove in post-processing. This is likely due to identifying lesions in the same way as the official evaluation code, which combines separate connected components into one lesion if they are still close enough to each other. Thus, the underlying ground truth still contained connected components of 50 voxels or fewer, so our model still learned to predict these structures.


Furthermore for Challenge 2, we investigated transfer learning approach to overcome the problem of small training data associated with the challenge. Freezing layers of Challenge 1's model, and continuing training on Challenge 2's dataset shows good results as seen in Table \ref{tab:frozen_layers}. 

We compared the results from freezing different layers against the performance of 2 baseline models: model trained on challenge 2 data only, and  model trained on challenge 1 data only. Surprising, model trained on Challenge 1 data only performed better on challenge 2 than the one trained challenge 2 data. Indicating the inadequacy of challenge 2 data, and the need to enhance challenge 2 dataset both in quantity and quality.

Overall we found that finetuning the model pretrained on challenge 1 data, lead to some improvements over the model trained only on challenge 1 data. Freezing the encoder led to decrease in performance over the challenge 1 model baseline, in terms of both Dice and HD95 scores. 
Freezing no layers and freezing the middle layers improved the Dice scores but HD95 saw a dip in performance. Freezing the decoder improved Dice scores across all tumor sub-regions and comparable HD95 scores (especially for WT). Looking ahead, we will be further experimenting with other frozen layer configurations, notably those focused on the decoder.

The strategies we implemented show improvements in tumor segmentation, however,  there is still a lot of room for improvement regarding better lesion-wise performance. 

We note that the state-of-the-art models, at least in adult glioma patients, achieve Dice scores close to $0.90$ or above, and we recognize that many more techniques can be implemented to push performance even higher, such as data augmentation, deep supervision, dropout, addition of extra input channels and complex loss functions that drive more focused learning. Further investigation will be done using such techniques to further optimize deep learning models in the realm of automatic brain tumor segmentation.

\section{Conclusion} 
Overall, our final models achieved fair performance in the BraTS 2023 validation phase, with the following lesion-wise Dice scores averaged across all three regions of interest: $0.79$ in Challenge 1; $0.72$ in Challenge 2; and $0.74$ in Challenge 3. We were able to thoroughly test and implement optimizing strategies for data pre-processing, prediction post-processing and transfer learning via the fine-tuning of models trained on one challenge to another challenge.



%
%
%
\bibliographystyle{splncs04}
\bibliography{bib}

\end{document}